\documentclass[sn-mathphys, Numbered, iicol]{sn-jnl}

\usepackage{comment}
\usepackage{graphicx}%
\usepackage{multirow}%
\usepackage{amsmath,amssymb,amsfonts}%
\usepackage{amsthm}%
\usepackage{mathrsfs}%
\usepackage[title]{appendix}%
\usepackage{xcolor}%
\usepackage{textcomp}%
\usepackage{manyfoot}%
\usepackage{booktabs}%
\usepackage{algorithm}%
\usepackage{algorithmicx}%
\usepackage{algpseudocode}%
\usepackage{listings}%

\theoremstyle{thmstyleone}%
\theoremstyle{thmstyletwo}%

\theoremstyle{thmstylethree}%

\begin{document}

\title[Article Title]{Optimization of geomagnetic shielding based on detection efficiency}


\author*[1,2,3]{\fnm{Sara R.} \sur{Cabo}}\email{rodriguezcsara@uniovi.es}

\author[4]{\fnm{Yasuhiro} \sur{Nishimura}}\email{nishimura@phys.keio.ac.jp}

\author[3,5]{\fnm{Sergio Luis} \sur{Suárez Gómez}}\email{suarezsergio@uniovi.es}

\author[1,2]{\fnm{Laura} \sur{Bonavera}}\email{bonaveralaura@uniovi.es}

\author[1,2]{\fnm{Maria Luisa} \sur{Sanchez}}\email{mlsr@uniovi.es}

\author[1,2]{\fnm{Jesús Daniel} \sur{Santos}}\email{jdsantos@uniovi.es}

\author[2,3]{\fnm{Francisco Javier} \sur{de Cos}}\email{fjcos@uniovi.es}

\affil[1]{\orgdiv{Departamento de Física}, \orgname{Universidad de Oviedo}, \orgaddress{\street{Calvo Sotelo 18}, \city{Oviedo}, \postcode{33007}, \state{Asturias}, \country{España}}}

\affil[2]{\orgdiv{MOMA}, \orgname{Instituto Universitario de Ciencias y Tecnologías Espaciales de Asturias (ICTEA)}, \orgaddress{\street{Independencia 13}, \city{Oviedo}, \postcode{33004}, \state{Asturias}, \country{España}}}

\affil[3]{\orgdiv{Departamento de Explotación y prospección de Minas}, \orgname{Universidad de Oviedo}, \orgaddress{\street{Independencia 13}, \city{Oviedo}, \postcode{33004}, \state{Asturias}, \country{España}}}

\affil[4]{\orgdiv{Departament of Physics}, \orgname{University of Keio}, \orgaddress{\street{Hiyoshi 3-14-1, Yokohama 223-8522, Japan}}}

\affil[5]{\orgdiv{Departamento de Matemáticas}, \orgname{Universidad de Oviedo}, \orgaddress{\street{Calvo Sotelo 18}, \city{Oviedo}, \postcode{33007}, \state{Asturias}, \country{España}}}


\abstract{Due to the progressive increase in size of the latest Cherenkov-type detectors, it is becoming increasingly important to design a suitable compensation system based on spins of the Earth's magnetic field to ensure the correct operation of the photomultipliers (PMTs). Until now, most studies have assessed the correct functioning of such a system by the proportion of PMTs experiencing more than 100 mG of magnetic field perpendicular to their axis. In the present study, we discuss whether this evaluation parameter is the most appropriate and propose the average residual perpendicular magnetic field $<B_{perp}>$ as an alternative that more closely reflects the loss of detection efficiency of PMTs. A compensation system design is also proposed that offers good results as well as being economical to optimise this parameter.}




\maketitle

\section{Introduction}\label{sec1}

The study of neutrinos is one of the cornerstones of modern particle physics. Their unique properties make them one of the keys to understand matter at its most fundamental level, as well as the fundamental forces that govern the universe. The discovery of neutrino oscillations \cite{Kajita} and their extremely low mass are unique properties that remain to be fully characterized and understood. Additionally, neutrinos are linked to astrophysical phenomena such as supernovae \cite{Supernova} and could prove useful in the search for dark matter characterization or proton decay \cite{KAbe}. Comprehensive study of neutrinos not only deepens our understanding of these particles but also sheds light on various universal phenomena.\\

For these reasons, different approaches to detect such particles are being developed worldwide. Among them, one of the most widely used is the Cherenkov detector. It operates by detecting photons that constitute Cherenkov light, produced when charged particles, generated by weak interactions between neutrinos and atoms, travel faster than light through a dielectric medium \cite{Cherenkov}. While water is the most commonly used dielectric medium, alternatives such as liquid argon, as in the DUNE detector, are also being explored \cite{DUNE}.\\ 
Notable examples of experiments leveraging this phenomenon include Super-Kamiokande \cite{SuperK}, Hyper-Kamiokande \cite{KAbe}, DUNE \cite{DUNE}, JUNO \cite{JUNO} or NOvA \cite{NOVA}.\\

A significant challenge for these detectors is the loss of detection efficiency due to the Earth's magnetic field. In these experiments, photomultiplier tubes (PMTs) covering the detector's entire inner surface collect the Cherenkov light produced by neutrino interactions. The PMTs convert incident photons into electrical signals, allowing for the reconstruction of interaction characteristics. The geomagnetic field deflects the electrons' trajectory within the PMTs, preventing them from properly reaching the anode and thus reducing detection efficiency \cite{PMT}.\\
Therefore, a geomagnetic field compensation system is crucial for optimal detector performance and accurate data collection. Some detectors, like Kamiokande, address this issue by encasing each PMT in a $\mu$-metal mesh \cite{Mumetal}. However, despite it provides an effective passive shielding widely used in physics experiments, it also has some disadvantages such as the fact that it can be a source of radioactive contamination \cite{BOREXINO}. It is also static and cannot dynamically adjust to changing magnetic field conditions and, despite its effectiveness, $\mu$-metal may leave residual magnetic fields that can still influence the PMT's performance \cite{Tomiya}.\\
Although PMTs remain the predominant method for photon detection in these experiments, silicon photomultipliers (SiPMs) are being considered as a potential alternative. SiPMs are compact, insensitive to magnetic fields, and free from issues like electron deflection. However, their adoption in large-scale neutrino experiments is still limited due to factors such as cost, scalability, and a relatively lower dynamic range compared to PMTs \cite{SiPM}. As such, while SiPMs show promise for future experiments, PMTs remain the standard choice for now.\\

An increasingly adopted strategy, especially when wanting to compensate magnetic field in large volumes, involves designing a coil system on the detector's inner surface, behind the PMTs and aligned with the geomagnetic field. When energized, these coils generate a magnetic field of comparable strength but opposite direction to the Earth's field. Developing an appropriate design, compensation coils can reduce the residual magnetic field to a much lower level compared to $\mu$-metal, helping to minimize PMT performance degradation \cite{Tomiya}. In addition, the current flowing through the coils can be calibrated, offering flexibility as magnetic conditions change.\\
However, this approach can also have some drawbacks, such as the fact that achieving the design of the coil system is not always straightforward and can sometimes be overly complex \cite{BOREXINO}, \cite{Mipaper}. It also requires an active power source. Besides, installing and maintaining compensation coils can be expensive and require significant space, and the presence of such large coils can make it difficult to access the detector. Nevertheless, it is a preferable and more convenient alternative in some experiments and is becoming increasingly important in the development of modern particle detectors.\\

Despite its crucial role in detector operation, this compensation system has received limited attention in literature. Understanding its design factors is complex, with potential implications for both experimental costs and detection efficiency, as demonstrated in \cite{Mipaper} and \cite{JUNOmag}.\\

The study in \cite{Mipaper} examines how factors like coil spacing, PMT-to-turn distance, detector size, and geomagnetic field strength affect compensation. It evaluates compensation effectiveness based on the proportion of PMTs experiencing over 100 mG of residual perpendicular magnetic field. This paper examines the suitability of this evaluation parameter and proposes alternative metrics that may more accurately reflect detection efficiency loss and potentially streamline the coil system design. Following \cite{Mipaper}, this study focuses on cylindrical detectors, a common geometry in current designs, which present more complex compensation challenges compared to spherically symmetric detectors.\\

The first section outlines the study objectives, experimental design, and simulation methodology. The subsequent section provides a detailed analysis of how various evaluation parameters reflect efficiency losses in a simulated coil-based compensation system. Additionally, we propose an alternative system design that is more cost-effective and simpler to install, while maintaining comparable performance. The final section summarizes the study's key findings and conclusions.\\

\section{Objectives and experiment design}\label{sec2}

\subsection{Objectives}

PMTs lose detection efficiency when exposed to a magnetic field perpendicular to their axis, as the field deflects the electron path toward the anode. In prior analyses of coil-based compensation system designs \cite{KAbe}, \cite{Mipaper},  \cite{SuperKcomp}, the proportion of PMTs exposed to more than 100 mG of residual perpendicular magnetic field after compensation was used as a benchmark to evaluate the system’s performance. This criterion was chosen based on studies showing that above 100 mG, the efficiency loss becomes noticeable and exceeds 1\% \cite{KAbe}, \cite{umbral1}, \cite{umbral2}.\\

However, as will be demonstrated in later sections, a detailed analysis of the relationship between PMT detection efficiency loss and the magnetic field reveals a more complex interaction.\\
This paper aims to provide a more detailed characterization of this relationship and to evaluate the suitability of using the proportion of PMTs exposed to more than 100 mG of perpendicular magnetic field as a metric for geomagnetic field compensation. Alternative parameters, such as the average perpendicular magnetic $<B_{perp}>$ or the average total magnetic field $<B_{total}>$, are proposed as more effective metrics for characterizing and comparing compensation coil system designs. In addition, depending on the parameter to be optimized based on the detector's compensation objectives, coil system design alternatives are proposed that provide optimum results while being more economical and easier to install.\\

\subsection{Design of the system}

The detector geometry is assumed to be cylindrical, with its center as the origin of the coordinate system used throughout this study, as shown in Figure 1. In this reference system, the Z-axis is vertical, the X- and Y-axes are horizontal, and the geomagnetic field at any point within the detector is represented as a three-dimensional vector.\\

\begin{figure}[!t]
\centering
\includegraphics[width=2.5in]{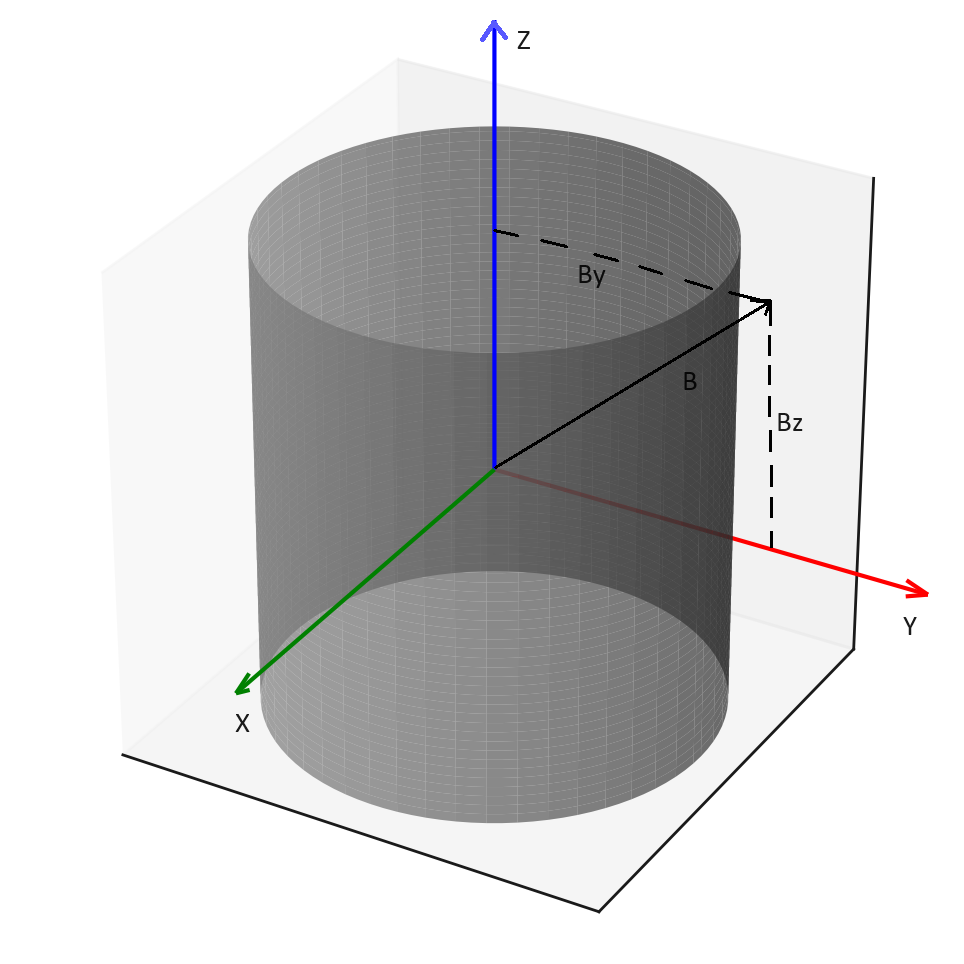}
\caption{Schematic representation of the generic cylindrical detector under consideration. The reference system used is also shown, with origin at the centre of the detector, as well as an example of a geomagnetic field vector.}
\label{Figure1}
\end{figure}

The compensation strategy involves placing coils on the detector’s inner surface, aligned with the geomagnetic field components. When an optimized current flows through these coils, it generates a magnetic field in the opposite direction, effectively canceling the Earth’s magnetic field.\\
However, it is important to note that the magnetic field generated by the coils is not uniform across the detector volume. Due to factors such as the finite size of the coils, edge effects, and the geometry of the system, the magnetic field distribution varies spatially within the detector. Therefore, the residual magnetic field at each PMT is position-dependent,the optimized current intensities correspond to minimizing the overall effect of the residual magnetic field, as represented by the chosen parameters (e.g., $P_{100}$, $<B_{perp}>$, $<B_{total}>$) over the entire detector., rather than to a single uniform field value across the entire detector.\\

\begin{figure}[!t]
\centering
\includegraphics[width=2.5in]{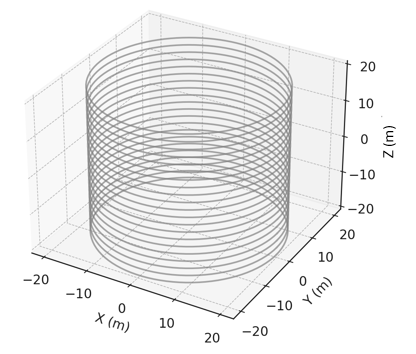}
\caption{Simulation of circular coil system in red centred on Z-axis for geomagnetic field compensation in a tank of 40 m height and diameter.}
\label{Figure2}
\end{figure}

\begin{figure}[!t]
\centering
\includegraphics[width=2.5in]{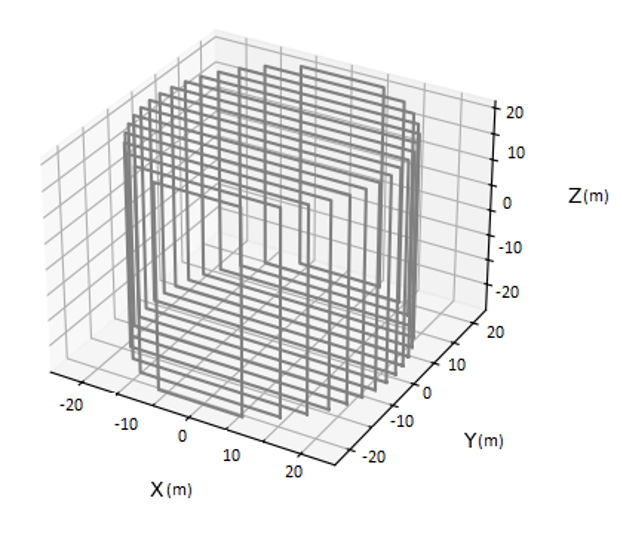}
\caption{Simulation of rectangular coil system in red centred on Y-axis for geomagnetic field compensation in a tank of 40 m height and diameter.}
\label{Figure3}
\end{figure}

In order to avoid having to install three sets of coils, one for each component of the geomagnetic field, which would make the installation too complex and costly, the X component of the field is mathematically set to zero.\\
This is achieved not by physically rotating the detector but by defining a reference coordinate system centered on the detector and rotating this reference system in the XY plane. Such a transformation ensures that the X component of the geomagnetic field aligns with zero in the chosen coordinates.\\
In this way, the compensation system is simplified to require only two sets of coils. Given the geometry of the detector, these two sets of coils will be circular coils centred on the Z axis to compensate for the vertical component $B_{Z}$ of the field, as shown in figure \ref{Figure2}; and rectangular coils centred on the Y axis to compensate for the non-zero horizontal component $B_{Y}$, as shown in figure \ref{Figure3}.\\

As mentioned, these coils are located on the inner surface of the detector, which means that the circular coils have the same radius as the tank and are arranged from the top and bottom. As for the rectangular coils, they are circumscribed, so that their height is also that of the detector, but the length of their horizontal sides varies according to the radius. For this study, normal conductive (non-superconducting) coils were assumed. On the other hand, the optimal number of coils for each set depends on the specific dimensions of the detector, as thoroughly analyzed in \cite{Mipaper}.\\

Furthermore, in order to calculate the residual magnetic field in each of the PMTs, which is the objective, it is necessary to determine their exact position. In this case, photomultiplier tubes (PMTs) with a diameter of 50 cm were used, taking as a reference advanced models specifically designed for the study of high-energy physics in large-scale Cherenkov detectors \cite{EleccionPMTs}, such as the Hamamatsu R12860 and Hamamatsu R3600 used in the Super-Kamiokande neutrino detector \cite{SuperKcomp}.\\
The distance between the centres of adjacent PMTs has been set at 70 cm. The PMTs are mounted on frames inside the detector, forming a second inner cylinder. It is important to clarify that the size and position of the coils are fixed along the inner surface of the tank. Thus, changing the distance between the PMTs and the coils corresponds to moving the PMTs closer to or farther from the tank walls, which effectively changes the size of the inner detection cylinder. The larger the distance from the tank walls, the smaller the effective detection volume. The free spaces between adjacent PMTs are covered with opaque black foils to absorb the incident radiation, although they do not influence the magnetic field simulation and will be treated as hollow spaces in it \cite{KAbe}, \cite{SuperKcomp}.\\

The simulation assumes that the geomagnetic field is well-characterized at the detector’s location based on a previous detailed study of the magnetic conditions at the placement site, providing a precise understanding of its direction and magnitude. While the coils are assumed to be aligned accordingly, certain practical factors could introduce small deviations in the alignment of the geomagnetic field relative to the compensation system. For instance, local magnetic anomalies, installation errors, or fluctuations in the geomagnetic field could cause slight misalignment. These variations, although small, might influence the simulation results. However, the overall trends and conclusions drawn from the simulation remain valid for the scope of the study.\\

\subsection{Calculation of total and perpendicular magnetic field}

The total residual magnetic field to which the PMTs are exposed when current flows through the windings can be easily calculated. This will be referred to as $B_{total} = (\Delta B_{X}, \Delta B_{Y}, \Delta B_{Z})$ this residual magnetic field, where  $B_{total} = (B_{X_{geo}}+B_{X_{coils}}, B_{Y_{geo}}+B_{Y_{coils}}, B_{Z_{geo}}+B_{Z_{coils}})$, i.e. the total residual field over each PMT is the sum of the geomagnetic field and the magnetic field generated by the coils at that exact position. The magnitude of this total residual magnetic field, which is used as a parameter to evaluate the compensation results, is calculated as the modulus of any three-dimensional vector.
\begin{equation}
    B_{total} = \sqrt{\Delta B_{X}^{2}+\Delta B_{Y}^{2}+\Delta B_{Z}^{2}}
\end{equation}

It is important to note that the geomagnetic field has a zero $B_{X_{geo}}$ component as the reference frame has been introduced. Both circular and rectangular coils produce a magnetic field in this direction, which must be considered in the calculations. However, simulations indicate that the magnetic field in this component is an order of magnitude smaller than that produced by the coils in the other two directions, as they are not specifically centered on this axis. While this influence must be considered, it is not expected to significantly affect the geomagnetic field compensation.\\

Another crucial parameter for assessing the effectiveness of compensation is the residual magnetic field perpendicular to the PMTs, denoted as $B_{perp}$. PMTs experience the greatest loss in detection efficiency when exposed to magnetic fields perpendicular to their axis, as these fields deflect the electron trajectories. $B_{perp}$ quantifies the magnetic field component responsible for efficiency decrease, while disregarding the parallel component, which requires significantly higher magnitudes to impact performance.\\

\begin{figure}[!t]
\centering
\includegraphics[width=2.5in]{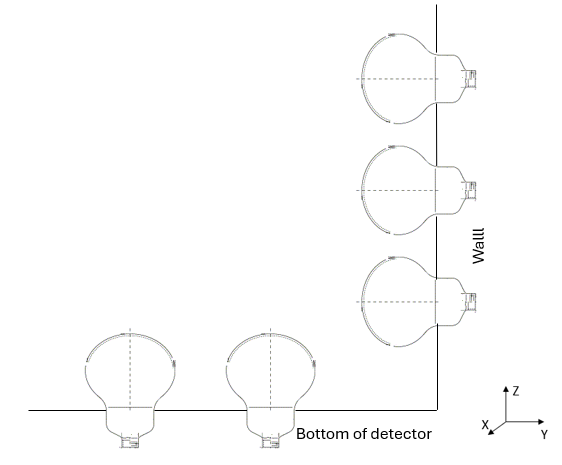}
\caption{Orientation of PMTs in a cylindrical detector. The orientation depends on whether they are on the lids or on the walls. The XYZ system of reference is also depicted.}
\label{Figure4}
\end{figure}

Calculating the perpendicular magnetic field is more complex due to the varying orientation of PMTs relative to the reference system, which depends on their position in the detector. Figure \ref{Figure4} illustrates the orientation of PMTs on the walls and lids. 

\begin{figure}[!t]
\centering
\includegraphics[width=2.5in]{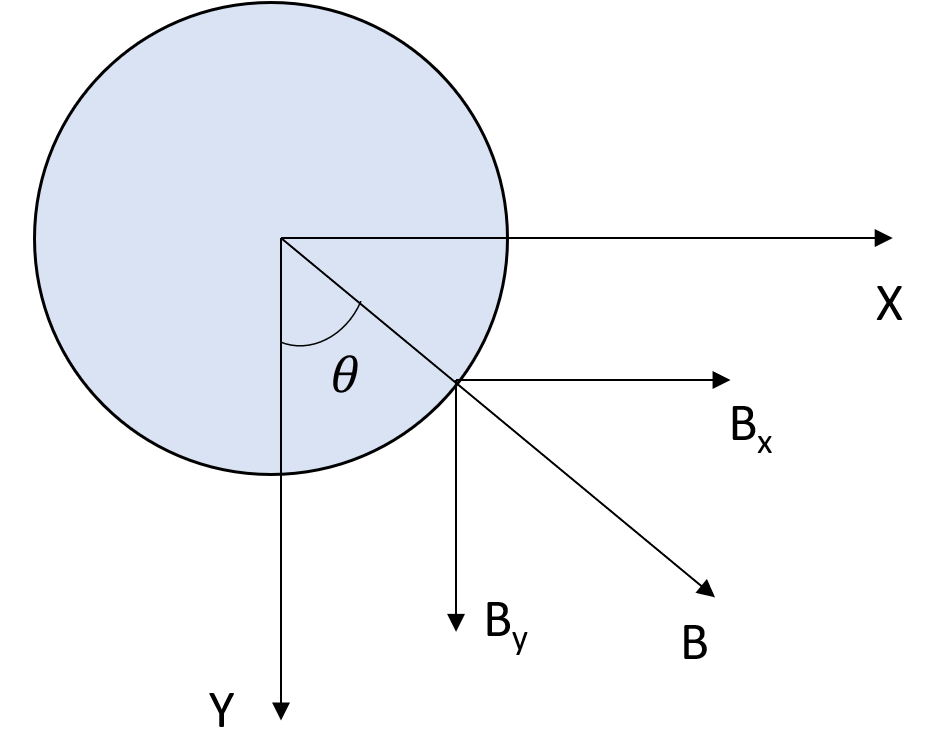}
\caption{Decomposition of the magnetic field at a generic point on the surface of the detector into its various components.}
\label{Figure5}
\end{figure}

We derive separate expressions for Bperp: one for PMTs on the side walls and another for those on the top and bottom lids. Equation \ref{Campo paredes} presents the expression for $\Delta B_{perp}$ for the PMTs on the walls.
\begin{equation}
\label{Campo paredes}
B_{perp}^{walls} = \sqrt{(\Delta B_{X}\cdot sin\theta + \Delta B_{Y}\cdot cos\theta)^{2}+\Delta B_{Z}^{2} }
\end{equation}

where $\theta$ represents the angle between each PMT's axis and the Y-coordinate axis. Figure \ref{Figure5} illustrates the decomposition of the horizontal magnetic field component for wall-mounted PMTs as a function of angle $\theta$.\\ 

This calculation considers only the components of $\Delta B_{X}$ and $\Delta B_{Y}$ perpendicular to the PMT axis, while the entire $\Delta B_{Z}$ component contributes as it is always perpendicular to the horizontally placed wall PMTs. The expression for the magnetic field perpendicular to the lid PMTs, shown in equation \ref{Campo tapas}, is simpler due to their parallel orientation to the Z axis. Consequently, the Z component of the magnetic field is disregarded, while the X and Y components contribute fully.”

\begin{equation}
\label{Campo tapas}
B_{perp}^{lids} = \sqrt{\Delta B_{X}^{2} + \Delta B_{Y}^{2} }
\end{equation}

\

To calculate the magnetic field created by each coil at the exact position of each PMT, the Biot-Savart law is applied within the framework of Maxwell's classical theory of electromagnetism.\\

\subsection{Calculation of efficiency}\label{sec3}

Detection efficiency of a PMT is defined as the ratio of photons detected to photons incident on the photocathode. It is the product of Quantum Efficiency (QE), the fraction of incident photons converted into photoelectrons, and Collection Efficiency (CE), the fraction of emitted photoelectrons that contribute to the final signal.\\

To experimentally measure this efficiency, a controlled light source is used to emit photons towards the PMT, while varying magnetic field strengths and directions are applied using Helmholtz coils. A laser diode (LD) provides stable light intensity to ensure consistent photon emission, while the PMT's response is monitored under different conditions, such as orientation and magnetic field exposure.\\
In these measurements, the magnetic field’s influence on the collection of photoelectrons is studied by varying the current in the compensation coils, adjusting the magnetic field's strength and direction. The PMT is placed at the center of the coils, and its orientation is adjusted to apply magnetic fields along different axes. This setup allows to quantify how the magnetic field affects both Quantum and Collection Efficiencies by comparing the detected signal under varying conditions. The voltage applied to the PMTs is also varied, though for this study, the lowest voltage (1600 V) is primarily considered to maximize the observed effect of the magnetic field on the efficiency. All the detailed information about the experimental procedure and results can be found in \cite{Tomiya}.\\

Data from the experiments is analyzed to determine the relationship between the applied magnetic field and PMT efficiency. This provides a detailed understanding of how magnetic fields influence PMT performance, particularly the electron trajectories inside the device. An accurate interpolation function is then derived to account for efficiency losses based on field strength and PMT voltage, serving as the basis for the present study.\\

\subsection{Design of the experiment}

The current literature on geomagnetic field compensation systems for Cherenkov-type detectors lacks consensus on the most suitable parameter to evaluate system performance. Some studies use the proportion of PMTs exposed to more than 100 mG of residual perpendicular magnetic field \cite{KAbe}, \cite{SuperKcomp} while others focus on the residual total magnetic field $<B_{total}>$ \cite{JUNOmag}.\\

To identify a suitable parameter and analyze the pros and cons of each, this study conducts simulations comparing different parameters based on their impact on PMT detection efficiency. Specifically, three parameters are compared: the proportion of PMTs exposed to more than 100 mG of perpendicular magnetic field, the average perpendicular magnetic field $<B_{perp}>$ and the average total residual magnetic field $<B_{total}>$.\\

To highlight the differences between optimizing each parameter, simulations determine the optimal current required to minimize each one. These simulations show that minimizing one parameter may increase the others, requiring prioritization of which parameter to optimize.\\

Secondly, the behavior of $<B_{perp}>$ and $<B_{total}>$ will be examined, as was done in \cite{Mipaper} for the proportion of PMTs subjected to more than 100 mG of perpendicular field. This will be analyzed as a function of the distance between the PMTs and coils, as well as the distance between the coils. This relationship has proven critical for optimizing the compensation system.\\

The next step involves simulating the optimization of each of the three evaluation parameters for a given coil distance by varying the distance to the walls. Afterward, the PMT efficiency loss will be calculated for each case to determine which of the three parameters, when minimized, leads to a lower loss in detection efficiency. This will be done for different values of coil spacing to assess whether these quantities influence the choice of evaluation parameter.\\

Finally, the impact on design optimization of incorporating the elliptical coils proposed in \cite{Mipaper} which yielded excellent results in minimizing the proportion of PMTs subjected to excessive perpendicular magnetic fields, will be evaluated. This includes the addition of smaller-radius circular coils at both the top and bottom ends, considering the advantages and disadvantages of each in terms of installation.\\

The detector sizes considered in the experiment will be $h = D = 40\: m$ and $h = D = 60\: m$. The simulated geomagnetic field will have an intermediate value, corresponding to intermediate latitudes, with both components equal: $B_{geo} = 475\: mG$ with $B_{Y_{geo}} = B_{Z_{geo}} = 335.88\: mG$. The effect of detector size, detector ratios, and different geomagnetic field values on compensation has been studied in detail in \cite{Mipaper}.\\

\subsection{Simulation}

The program used to simulate geomagnetic field compensation in the detector consists of three Python modules. The first module defines the geometry of the coils, which, in this case, are circular, rectangular, and, in some axes, elliptical. The second module calculates the magnetic field generated by the coils, as defined in the previous module, at any given coordinate by directly applying Biot-Savart's law.\\

The main module imports the two previous modules and defines the characteristic parameters of the detector, such as its dimensions, as well as the exact position of the PMTs. This module has three functions. The central function defines the position, geometry, size, and current intensity of the coils. It then calculates the total magnetic field, which is the sum of the geomagnetic field and the magnetic field generated by all the coils, broken down by component, as well as the magnetic field perpendicular to the axis of each PMT. The function returns the average total and perpendicular magnetic fields across the detector, as well as the percentage of photomultipliers exceeding the 100 mG perpendicular magnetic field limit.\\

The second function produces a histogram of the perpendicular magnetic field and a graphical representation of the cylindrical tank, highlighting the positions of photomultipliers exceeding 100 mG of residual perpendicular magnetic field.\\
The third function implements a genetic algorithm to optimize the current intensity applied to the coils.\\

A genetic algorithm is a programming technique inspired by genetic processes and natural selection, used for the optimisation of non-linear multivariate problems. \cite{Genetico}. The algorithm starts with a population of N individuals, each represented by a chromosome. Each chromosome consists of a number of genes corresponding to the values of the parameters to be optimised. Thus, to optimise $n$ parameters, each chromosome contains $n$ values, one for each parameter. In this application, the genes correspond to the current intensity through the circular and rectangular coils. For simplicity and cost-efficiency, it is assumed that the same intensity flows through all coils of the same geometry.\\
Another key element is defining a fitness function that takes these chromosomes as input and returns a measure of their effectiveness as solutions to the problem. In this case, the fitness function will be the program used so far to calculate the residual perpendicular magnetic field over the photomultipliers (PMTs), returning the number of PMTs exposed to less than 100 mG. The aim is to maximise this number of PMTs.\\

The algorithm identifies the best individuals based on a fitness function, prioritizing gene values that maximize the number of PMTs exposed to a magnetic field lower than 100 mG. It then replaces the least fit individuals with combinations of the best ones, introducing random mutations as the program iterates to find the optimal configuration of the system. The greater the number of genes or parameters to be optimized and the wider the range of possible values for each, the more iterations are required to achieve convergence.\\

Finally, a separate program calculates the efficiency loss of each PMT using the experimentally obtained interpolation function for a given voltage, with the total magnetic field and its components at the position of each PMT as input.\\

\section{Results}\label{sec4}

After establishing the design of the coil-based compensation system for the detector by setting the distance between the coils, it is crucial to note that the optimal current intensity for effective compensation varies significantly depending on the evaluation parameter used.\\
In the absence of coils, the geomagnetic field is assumed to be constant throughout the detector, with a total magnitude of $B=475\: mG$ and components $B_{Y}=B_{Z}=335.88\: mG$. Under these conditions, the values of the evaluation parameters would remain constant: all PMTs would be above the 100 mG threshold, $<B_{perp}>\: = 384.54\pm 52.91\: mG$, and $<B_{total}>\: = 475\: mG$. Consequently, the average detection efficiency loss would be $27.02\pm 5.19\%$, with losses ranging between 15\% and 36\% across all PMTs. These results demonstrate the significant efficiency reduction caused by the unmitigated geomagnetic field and highlight the importance of an effective compensation system.\\

\begin{figure}[!t]
\centering
\includegraphics[width=2.5in]{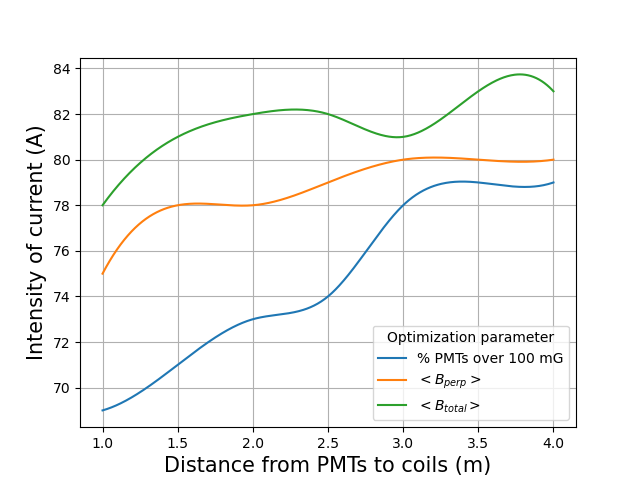}
\caption{Optimal intensity of current of rectangular coils optimizing $P_{100}$, $<B_{perp}>$ and $<B_{total}>$ for different values of the distance between PMTs and coils and a distance between coils set to 2 m.}
\label{Figure6}
\end{figure}

\begin{figure}[!t]
\centering
\includegraphics[width=2.5in]{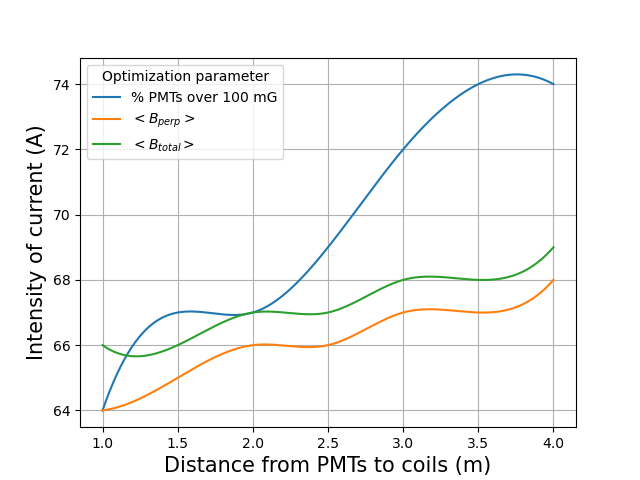}
\caption{Optimal intensity of current of circular coils optimizing $P_{100}$, $<B_{perp}>$ and $<B_{total}>$ for different values of the distance between PMTs and coils and a distance between coils set to 2 m.}
\label{Figure7}
\end{figure}

\begin{figure}[!t]
\centering
\includegraphics[width=2.5in]{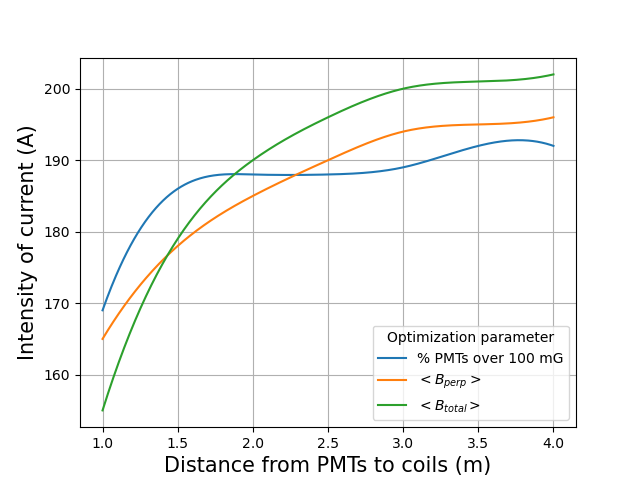}
\caption{Optimal intensity of current of rectangular coils optimizing $P_{100}$, $<B_{perp}>$ and $<B_{total}>$ for different values of the distance between PMTs and coils and a distance between coils set to 5 m.}
\label{Figure8}
\end{figure}

\begin{figure}[!t]
\centering
\includegraphics[width=2.5in]{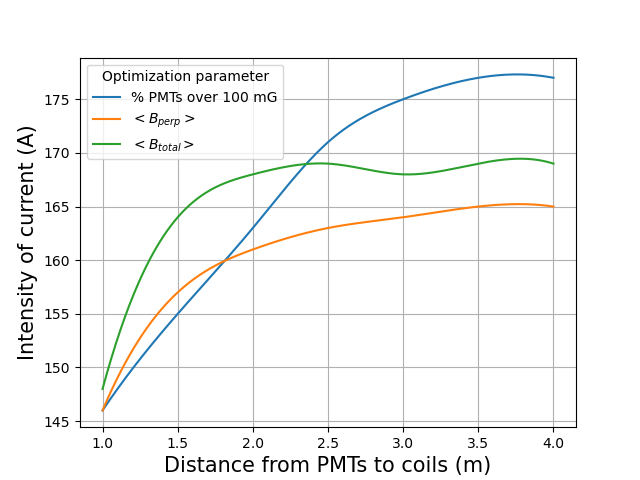}
\caption{Optimal intensity of current of circular coils optimizing $P_{100}$, $<B_{perp}>$ and $<B_{total}>$ for different values of the distance between PMTs and coils and a distance between coils set to 5 m.}
\label{Figure9}
\end{figure}

While one might assume that minimizing one of the three parameters considered in this study—the proportion of PMTs exposed to more than 100 mG of perpendicular magnetic field ($P_{100}$), the average perpendicular field $<B_{perp}>$, and the average total magnetic field $<B_{total}>$—would inherently minimize the other two, this is not the case. Each parameter exhibits a distinct optimal current value, and any deviation from this value results in an increase in the corresponding parameter. To illustrate this phenomenon, Figures \ref{Figure6}, \ref{Figure7}, \ref{Figure8}, and \ref{Figure9} depict the optimal current intensity for both circular and rectangular coils associated with each evaluation parameter as a function of the distance between PMTs and walls, with fixed inter-coil distances of $d_{C} = 2\: m$ and $d_{C} = 5\: m$.

Figures \ref{Figure6} and \ref{Figure7} illustrate the behavior of the optimal current intensity for each parameter for rectangular and circular coils, respectively, with a coil spacing of 2 m. In the case of rectangular coils, a significant difference in current intensities is observed, reaching almost 10 A between the optimal current for $P_{100}$ and that for $<B_{total}>$. The latter consistently exhibits the highest current intensity. The current for $P_{100}$ remains at the lowest levels, irrespective of the distance between PMTs and coils. While the current for $<B_{total}>$ is always the highest, the difference between the two currents decreases with increasing distance.\\
Circular coils, however, exhibit a distinct behavior. The difference between the optimal intensities for $<B_{perp}>$ and $<B_{total}>$ remains relatively constant throughout the considered range. Contrary to rectangular coils, where the optimal intensity for $P_{100}$ is consistently the lowest, circular coils demonstrate the opposite trend. The optimal current for $P_{100}$ becomes the highest for a significant portion of the studied distance range between PMTs and coils. This difference becomes particularly pronounced when the distance exceeds 2 m.\\
Intersection points are observed, representing current values that are optimal for two parameters simultaneously at a specific distance between PMTs and coils. For example, at a distance $d_{W} = 2\: m$, a current of $I_{r} = 67\: A$ minimizes both the proportion of PMTs with excess field and the average total magnetic field $<B_{total}>$. However, this current would be too high for optimizing the average perpendicular magnetic field component $<B_{perp}>$.\\

The bends and valleys observed in the optimal current intensity curves can be attributed to a combination of factors. First, the non-uniform magnetic field generated by the finite-sized compensation coils may exhibit nonlinear behavior, especially at certain distances from the PMTs, due to edge effects or interference between field components. Second, the geometry of the magnetic field plays a role: while $<B_{total}>$ reflects all field components, $<B_{perp}>$ considers only the perpendicular contribution, leading to potential cancellations or reductions at specific distances. Additionally, the influence of the coils is distance-dependent but not perfectly linear, as the magnetic field contributions from different coil segments may interfere constructively or destructively.\\

Figures \ref{Figure8} and \ref{Figure9} demonstrate the impact of increasing the inter-coil distance ($d_{C}$) to 5 m on the optimal current intensities. With this increased spacing, more intersection points emerge between the different current intensities for rectangular coils, and the overall difference between them becomes less significant. The optimal current for $P_{100}$ exhibits the highest values at lower $d_{W}$ values and the lowest values at higher $d_{W}$ values. Conversely, the optimal current for $<B_{total}>$ displays the opposite trend. In the case of circular coils, the behavior resembles that observed in Figure \ref{Figure7}, with the difference between the optimal currents for $<B_{perp}>$ and $<B_{total}>$ remaining consistent throughout the range. The optimal current for $P_{100}$ progressively increases, eventually surpassing the other two, though the difference is less pronounced compared to the previous case.\\

\begin{figure}[!t]
\centering
\includegraphics[width=2.5in]{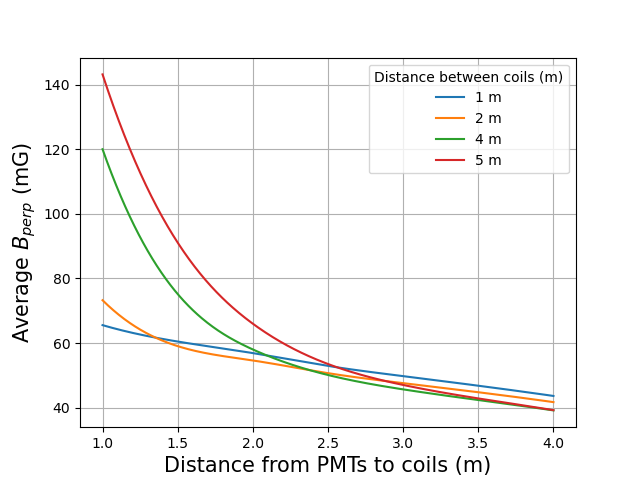}
\caption{Average $B_{perp}$ for different values of the distance to the side walls and the caps in a detector of $h = D = 40\: m$.}
\label{Figure10}
\end{figure}

\begin{figure}[!t]
\centering
\includegraphics[width=2.5in]{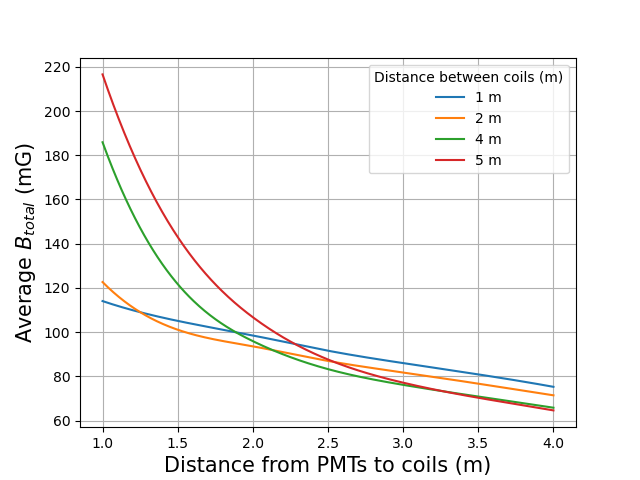}
\caption{Average $B_{total}$ for different values of the distance to the side walls and the caps in a detector of $h = D = 40\: m$.}
\label{Figure11}
\end{figure}

These findings highlight the significant variation in optimal current intensities depending on the chosen evaluation parameter, as well as the distances between PMTs and walls and between coils. To effectively determine these distances, a detailed understanding of the behavior of the three field compensation evaluation parameters as a function of these distances is essential. While \cite{Mipaper} has previously investigated this relationship for $P_{100}$, Figures \ref{Figure10} and \ref{Figure11} present the same analysis for $<B_{perp}>$ and $<B_{total}>$. These figures depict the respective parameters as a function of the distance between PMTs and coils ($d_{W}$), considering various inter-coil distances ($d_{C}$). For both rectangular and circular coil sets, the inter-coil distance is assumed to be constant, with the optimal current value applied in each case.\\

The fundamental behavior of the parameters mirrors the observations in \cite{Mipaper} for the fraction of PMTs exposed to more than 100 mG. As the distance between PMTs and coils increases, the geomagnetic field compensation improves, irrespective of the inter-coil distance. Furthermore, the optimal inter-coil distance also increases with increasing $d_{W}$. The characteristic exponential increase in the compensation parameter for low $d_{W}$ values is observed for inter-coil distances greater than 4 m ($d_{C} > 4\: m$).\\

Notably, for a fixed distance between PMTs and coils ($d_{W}$), the optimal inter-coil distance may vary depending on the chosen evaluation parameter, as suggested by the differences in optimal current intensities. For instance, at $d_{W} = 2\: m$, the optimal inter-coil distance ($d_{C}$) for $P_{100}$ is 4 m, while for $<B_{perp}>$ and $<B_{total}>$, it is 2 m. Setting $d_{C}$ to 2 m, which is optimal for the latter two parameters, would lead to an increase in $P_{100}$ by almost 3 percentage points. Conversely, setting $d_{C}$ to 4 m would result in an increase of 3.4 mG for $<B_{total}>$ and 2.37 mG for $<B_{perp}>$.\\

\begin{figure}[!t]
\centering
\includegraphics[width=2.5in]{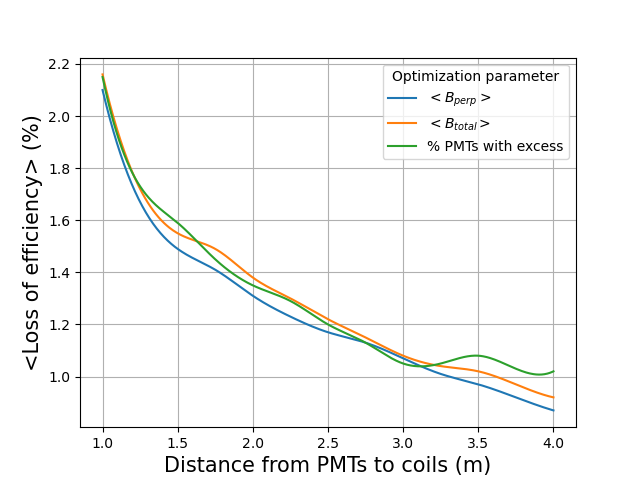}
\caption{Average loss of detection efficiency as a function of distance between PMTs and coils optimizing the three evaluation parameters considered and setting a distance between coils of 2 m and a detector size of $h = D = 40\: m$.}
\label{Figure12}
\end{figure}

\begin{figure}[!t]
\centering
\includegraphics[width=2.5in]{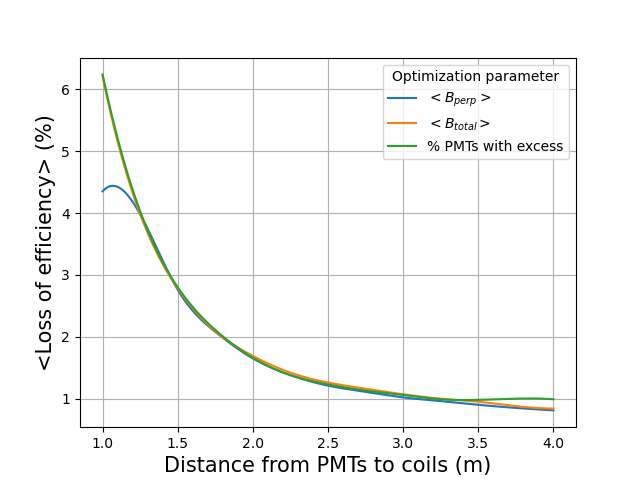}
\caption{Average loss of detection efficiency as a function of distance between PMTs and coils optimizing the three evaluation parameters considered and setting a distance between coils of 5 m and a detector size of $h = D = 40\: m$.}
\label{Figure13}
\end{figure}

Given that optimizing one compensation evaluation parameter can significantly impact others, understanding the influence of each parameter on PMT detection efficiency is crucial for establishing optimization targets and prioritizing parameters for minimization. To achieve this, the detection efficiency of all PMTs is calculated for a fixed inter-coil distance as a function of the distance between PMTs and walls. The current intensity is set to optimize each of the three considered evaluation parameters. Figures \ref{Figure12} and \ref{Figure13} present illustrative examples of the average detector efficiency loss as a function of the distance between PMTs and coils for inter-coil distances of 2 m and 5 m, respectively.\\

As evident from Figures \ref{Figure10} and \ref{Figure11}, and as expected, the average efficiency loss peaks at the minimum $d_{W}$ value due to improved geomagnetic field compensation at larger distances between PMTs and coils. With increasing distance, the efficiency loss progressively declines. For $d_{C}= 2\: m$, the maximum average loss is comparable across all evaluation parameters, averaging around 2.1\%. Notably, minimizing $<B_{perp}>$ yields the lowest average efficiency loss across most of the studied range of distances between PMTs and coils. The exception is a small interval around $d_{W} = 3\: m$, where optimizing $P_{100}$ results in slightly better performance. However, the maximum difference in average detection loss within this range, between optimizing one parameter or another, is marginal at only 0.04\%. At higher $d_{W}$ values, the difference in average efficiency loss becomes more pronounced, with $P_{100}$ leading to higher values. Furthermore, for intermediate distances between PMTs and coils ($d_{W} \in [1.6, 3.2]\: m$), minimizing $<B_{total}>$ results in a higher average efficiency loss.\\

When the inter-coil distance is set to a higher value, as depicted in Figure 13 with $d_{C} = 5\: m$, the average efficiency loss becomes more comparable across the three parameters throughout the entire dW range considered. The maximum average loss, again occurring at the lowest $d_{W}$ value, is significantly higher than in the previous case, reaching up to 6.24\% when minimizing $P_{100}$. However, this loss drops more rapidly, resulting in average efficiency loss values similar to those obtained with $d_{C} = 2\: m$ at the same distance between PMTs and coils for higher $d_{W}$ values. While the loss values are generally similar across the three evaluation parameters, minimizing $<B_{perp}>$ consistently yields the lowest average detection efficiency loss across the entire range. This difference is particularly pronounced at lower $d_{W}$ values, reaching almost 2\%. Similar to the previous case, the difference between parameters becomes more noticeable at higher distances between PMTs and walls, specifically for $d_{W} > 3.25\: m$. For intermediate distances ($d_{W} \in [2, 3.25]\: m$), minimizing $<B_{total}>$ results in the highest detection efficiency loss.\\

\begin{figure}[!t]
\centering
\includegraphics[width=2.5in]{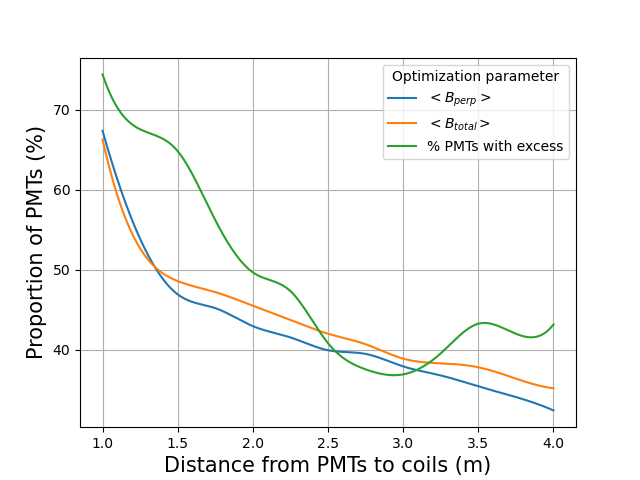}
\caption{Proportion of PMTs losing more than 1\% of detection efficiency as a function of $d_{W}$ optimizing the three evaluation parameters considered. Distance between coils is set to $d_{C} = 2\: m$ and the size of the detector simulated is of $h = D = 40\: m$.}
\label{Figure14}
\end{figure}

\begin{figure}[!t]
\centering
\includegraphics[width=2.5in]{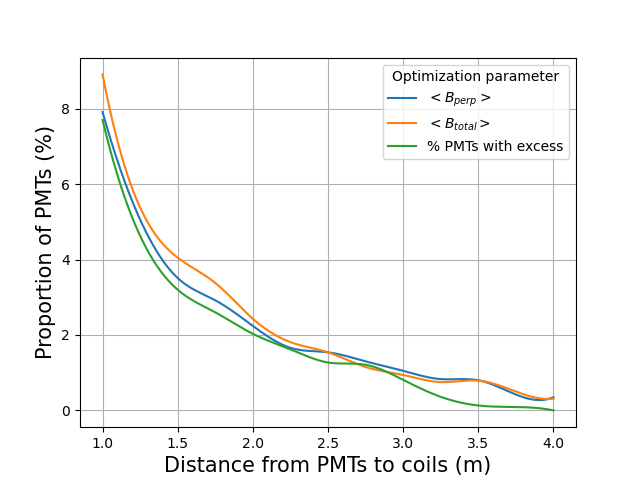}
\caption{Proportion of PMTs losing more than 5\% of detection efficiency as a function of $d_{W}$ optimizing the three evaluation parameters considered. Distance between coils is set to $d_{C} = 2\: m$ and the size of the detector simulated is of $h = D = 40\: m$.}
\label{Figure15}
\end{figure}

The average efficiency loss serves as a valuable metric to quantify the practical impact of minimizing the considered parameters on PMTs. However, to delve deeper into the effect of each parameter on efficiency loss, their influence on the number of PMTs experiencing a detection efficiency loss exceeding a certain threshold is also examined. Figures \ref{Figure14} and \ref{Figure15} illustrate the proportion of PMTs exhibiting a detection efficiency loss greater than 1\% and 5\%, respectively, as a function of the distance between PMTs and coils. This analysis considers the optimization of each of the three evaluation parameters while maintaining a fixed inter-coil distance of $d_{C} = 2\: m$.\\

The behavior observed in Figure \ref{Figure14} mirrors that of Figure \ref{Figure12} for the average efficiency loss at the same inter-coil distance, with the difference between optimizing one parameter versus another becoming more pronounced. As previously observed, larger distances between PMTs and coils result in lower efficiency losses. In this scenario, minimizing $<B_{total}>$ within the range of $d_{W} \in [1, 1.4]\: m$ yields slightly superior results compared to minimizing $<B_{perp}>$. However, for distances $d_{W} > 1.4\: m$, optimizing the latter parameter leads to lower average efficiency loss values.\\ 
Conversely, while optimizing $P_{100}$ generally produces the least favorable results across most of the range, the average efficiency loss reaches a minimum when this parameter is minimized at $d_{W} = 2.8\: m$. Notably, for $d_{W}$ values exceeding 2.8 m, the associated average efficiency loss increases once again, diverging significantly from the values obtained when optimizing the other two parameters. This discrepancy culminates in a maximum difference of 1.93\%.\\
The valleys and bends previously observed in earlier figures can also be seen in these graphics, again as a result of of the non-uniformity and non-linearity of the magnetic field, as well as the evaluation parameters.\\

Figure \ref{Figure15} presents the proportion of PMTs experiencing a significant detection efficiency loss exceeding 5\%. Notably, the difference between optimizing one parameter versus another is considerably smaller in this scenario compared to the previous analysis. However, a key distinction is that minimizing $P_{100}$ yields the most favorable results across nearly the entire range, except for $d_{W} \approx 2.75\: m$ where optimizing $<B_{total}>$ demonstrates marginally superior effectiveness (with a difference of only 0.07\%). This observation contrasts sharply with the previous analysis of PMTs losing more than 1\% efficiency, where optimizing $P_{100}$ consistently produced the least desirable outcomes.\\

\begin{figure}[!t]
\centering
\includegraphics[width=2.5in]{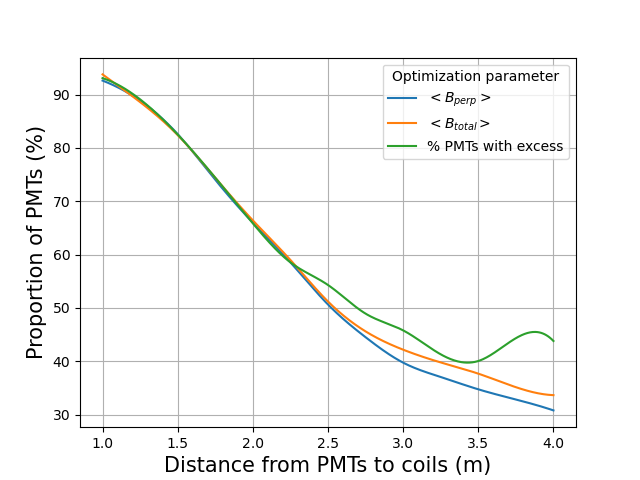}
\caption{Proportion of PMTs losing more than 1\% of detection efficiency as a function of $d_{W}$ optimizing the three evaluation parameters considered. Distance between coils is set to $d_{C} = 5\: m$ and the size of the detector simulated is of $h = D = 40\: m$.}
\label{Figure16}
\end{figure}

\begin{figure}[!t]
\centering
\includegraphics[width=2.5in]{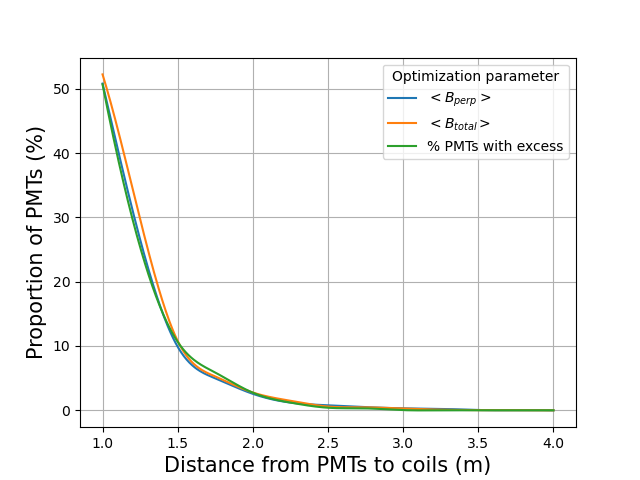}
\caption{Proportion of PMTs losing more than 5\% of detection efficiency as a function of $d_{W}$ optimizing the three evaluation parameters considered. Distance between coils is set to $d_{C} = 5\: m$ and the size of the detector simulated is of $h = D = 40\: m$.}
\label{Figure17}
\end{figure}

Figures \ref{Figure16} and \ref{Figure17} once again illustrate the number of PMTs exhibiting a detection efficiency loss exceeding 1\% and 5\%, respectively, as a function of the distance between PMTs and coils. However, in contrast to the previous analysis, the inter-coil distance is set to a higher value of $d_{C} = 5\: m$.\\

As anticipated from the results depicted in Figure \ref{Figure13}, the discrepancy in efficiency loss between optimizing different parameters diminishes as the inter-coil distance ($d_{C}$) increases. When analyzing losses exceeding 1\%, this difference is on the order of one hundredth for $d_{W} < 2.25\: m$. However, for larger $d_{W}$ values, the discrepancy becomes increasingly prominent. Optimizing yields the highest number of PMTs with less than 1\% detection efficiency loss, whereas the proportion of PMTs exposed to magnetic fields greater than 100 mG ($P_{100}$) results in the least favorable outcomes.\\ 
Conversely, when examining PMTs experiencing greater than 5\% detection efficiency loss, the variation based on the chosen evaluation parameter becomes negligible, particularly for $d_{W} > 2\: m$. Interestingly, in this scenario, minimizing P100 generally leads to fewer PMTs with greater than 5\% efficiency loss, with the exception of a narrow range ($d_{W} \in [1.5, 2]\: m$), where optimizing $<B_{total}>$ proves slightly more advantageous, albeit with a maximum difference of only 0.61\%.\\

The preceding results regarding detection efficiency yield several key conclusions. As evidenced by Figures \ref{Figure10} and \ref{Figure11}, both the average detection loss and the number of PMTs exceeding a specified detection efficiency loss threshold decrease as the distance between PMTs and coils ($d_{W}$) increases, a trend stemming from the increased irregularity of the magnetic field in closer proximity to the coils. Conversely, larger inter-coil distances ($d_{C}$) lead to higher efficiency losses for smaller $d_{W}$ values, although this effect rapidly diminishes with increasing $d_{W}$. Furthermore, increasing $d_{C}$ generally reduces the discrepancies observed between the outcomes of optimizing various parameters, with this difference becoming more apparent at higher $d_{W}$ values.\\

The parameter $<B_{perp}>$ proves to be the most effective for optimizing both the average loss of detection efficiency and the number of PMTs experiencing moderate efficiency losses across the entire range studied, with only negligible exceptions for specific $d_{W}$ values. However, when minimizing the number of PMTs suffering high efficiency losses, a different approach is more effective: minimizing the proportion of PMTs exposed to magnetic fields exceeding 100 mG yields significantly better results.\\

Although the proportion of PMTs exposed to magnetic fields exceeding 100 mG has traditionally served as a reference parameter for evaluating compensation (as this threshold was deemed significant for efficiency loss), the present findings demonstrate that magnetic fields below 100 mG, when perpendicular to the PMT axis, can still induce non-negligible efficiency losses. Consequently, this evaluation parameter proves valuable for minimizing the number of PMTs with substantial efficiency losses, but it may not be as representative when aiming to reduce losses across all levels. Additionally, while this parameter accurately reflects the number of PMTs experiencing significant losses, the difference observed between minimizing this parameter versus the other two is merely on the order of a tenth or even a hundredth for larger distances between PMTs and coils.\\

\subsection{Smaller circular coils}

\begin{figure}[!t]
\centering
\includegraphics[width=2.5in]{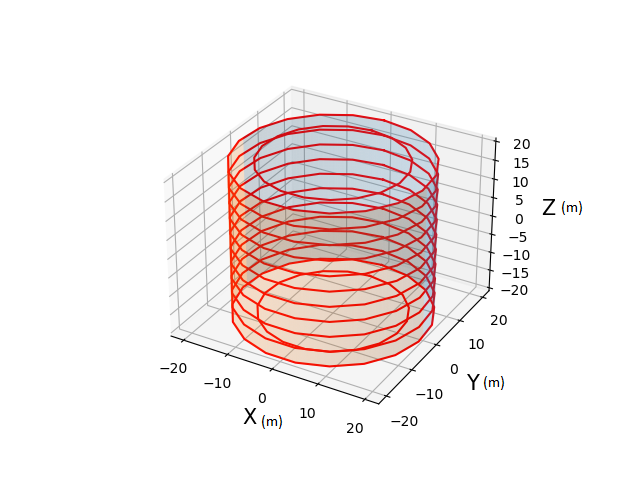}
\caption{Scheme of the circular coil assembly of a basic compensation system with smaller radius coils added at the top and bottom ends.}
\label{Figure18}
\end{figure}

In this subsection, a specific compensation scenario is investigated to compare the advantages and disadvantages of optimizing the proportion of PMTs with excess (as in previous studies \cite{KAbe},  \cite{Mipaper}, \cite{SuperKcomp}), versus optimizing $<B_{perp}>$ (as proposed in this article). Additionally, the incorporation of smaller-radius circular coils at the top and bottom ends of the detector (illustrated in Figure \ref{Figure18}) is proposed as a simpler and more cost-effective alternative to the elliptical coils introduced in \cite{Mipaper}. This approach aims to compensate for the geomagnetic field in the more challenging areas identified as the top and bottom edges of the walls.\\

For the present simulations, a medium-sized detector (h = D = 40 m) and a geomagnetic field corresponding to intermediate latitudes ($B_{geo} = 475\: mG$, $B_{Y_{geo}} = B_{Z{geo}} = 335.88\: mG$) are assumed. The distance between PMTs and walls ($d_{W}$) is set to 2 m, as a larger distance would result in overly simplistic compensation and hinder a robust comparison of the different scenarios analyzed herein. The inter-coil distance ($d_{C}$) is set to 2 m for both the minimization of $P_{100}$ and $<B_{perp}>$, as this value represents the optimal distance between turns for both cases, according to the findings from \cite{Mipaper} and Figure \ref{Figure10}, respectively.\\

Initially, geomagnetic field compensation is simulated by minimizing $P_{100}$. With both $d_{W}$ and $d_{C}$ fixed, the remaining step in completing this basic compensation system involves optimizing the current intensity flowing through the coils. This optimization yields a $P_{100}$ value of 9.71\% and an average residual perpendicular magnetic field ($<B_{perp}>$) of $57.4 \pm 29.76\: mG$.\\

Following the completion of the basic system, attention shifts to the most problematic areas, identified as the top and bottom edges of the walls (as observed in \cite{Mipaper}). To enhance current density in these regions, the number of turns in the circular coils at both ends is increased. Specifically, for minimizing $P_{100}$, this entails three turns on both the upper and lower coils. This modification reduces $P_{100}$ to 4.12\% and decreases to $53.23 \pm 22.84\: mG$. The average detection efficiency loss in this configuration is $1.17 \pm 0.87\%$.\\

\begin{figure}[!t]
\centering
\includegraphics[width=2.5in]{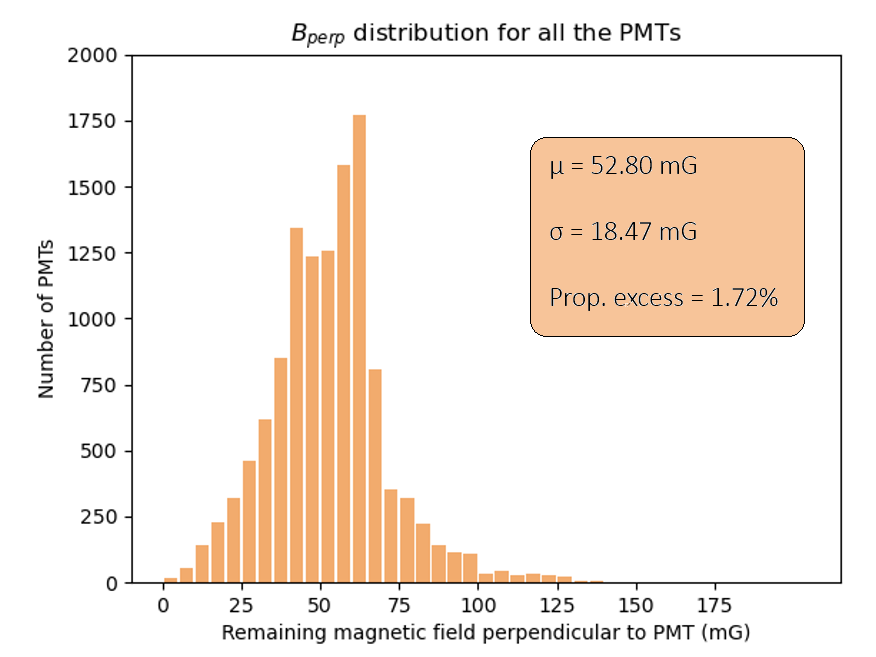}
\caption{Histogram of $<B_{perp}>$ after adding more turns to the top and bottom circular coils and elliptical coils to minimize the proportion of PMTs with excess.}
\label{Figure19}
\end{figure}

Building upon previous compensation efforts, the installation of elliptical coils with specific positions and dimensions on both lids (as proposed in \cite{Mipaper}) is simulated to further address the remaining uncompensated areas. These simulations demonstrate the effectiveness of elliptical coils, achieving a $P_{100}$ value of 1.72\% and a slight reduction in $<B_{perp}>$ to $52.8 \pm 18.47\: mG$ (Figure \ref{Figure19}). This configuration also leads to a decrease in the average efficiency loss to $1.13 \pm 0.64\%$, with 55.56\% of PMTs experiencing losses exceeding 1\% but none exceeding 5\%. \\

However, incorporating elliptical coils presents certain drawbacks. Their unique geometry, coupled with the requirement for precise coil dimensions to achieve optimal performance, may complicate installation within the detector. Moreover, unlike the basic coil system (where identical current flows through all coils of the same type, regardless of shape), each elliptical coil requires a different current intensity for optimal compensation. This can lead to high current demands (up to 135 A for the largest coils), potentially increasing the cost of power supplies.\\

To address these challenges, an alternative approach is explored in this work. Instead of elliptical coils, the impact of incorporating smaller-radius circular coils centered on both covers is simulated. To reduce the required number of power supplies, these coils are designed to utilize the same current as the other circular coils in the detector. Optimization efforts are then focused on adjusting the radii of these smaller coils to achieve the desired magnetic field compensation.\\

\begin{figure}[!t]
\centering
\includegraphics[width=2.5in]{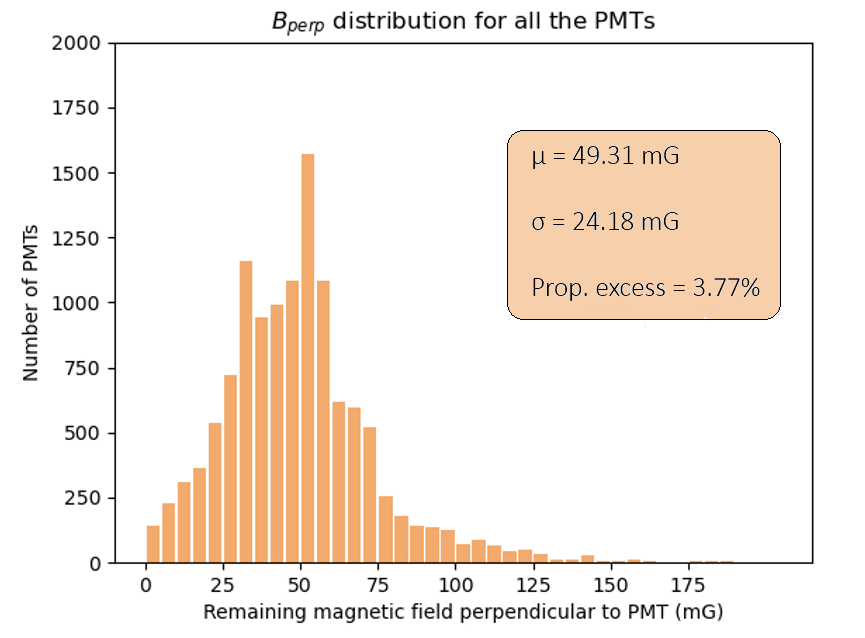}
\caption{Histogram of $<B_{perp}>$ after adding more turns to the top and bottom circular coils and circular coils of smaller radius to minimize the proportion of PMTs with excess.}
\label{Figure20}
\end{figure}

Optimizing the radii of both coils to minimize $P_{100}$ in this specific scenario yields a radius of R = 14 m for the top coil. Notably, the simulation reveals that adding a coil to the bottom cap does not further reduce $P_{100}$, regardless of its radius. The resulting configuration is depicted in Figure \ref{Figure20}.\\

In this case, a $P_{100}$ value of 3.77\% and an average $<B_{perp}>$ of $49.31 \pm 24.18\: mG$ are achieved. While incorporating this smaller-radius circular coil improves upon the configuration with only increased turns in the end coils, the resulting $P_{100}$ value remains nearly two percentage points higher than that achieved with elliptical coils. Conversely, $<B_{perp}>$ is reduced by almost four points compared to the elliptical coil configuration. Consequently, despite the higher $P_{100}$, the average efficiency loss decreases to 1.07\%, and the proportion of PMTs experiencing greater than 1\% loss falls to 44.68\%. However, the higher $P_{100}$ value leads to an increase in the proportion of PMTs with a substantial efficiency loss (greater than 5\%) to 0.62\%.\\

This example suggests that smaller-radius circular coils at both ends are less effective than elliptical coils for minimizing $P_{100}$. However, their efficacy in reducing $<B_{perp}>$ has been demonstrated. Therefore, the following example will compare the effectiveness of both approaches for minimizing $<B_{perp}>$, which has previously been shown to be more effective in reducing detection efficiency losses.\\

The aforementioned process is repeated, but this time the current values in the basic system are adjusted to minimize $<B_{perp}>$. This configuration yields a $P_{100}$ value of 12.12\% and a $<B_{perp}>$ of $54.62 \pm 34.44\: mG$. Optimizing the number of turns in the upper and lower circular coils results in four turns for the upper coil and three for the lower coil. This further reduces $P_{100}$ to 4.53\% and to $49.69 \pm 27.54\: mG$, resulting in an average efficiency loss of $1.12 \pm 1.05\%$. In this scenario, 41.94\% and 1.28\% of PMTs experience losses exceeding 1\% and 5\%, respectively.\\

\begin{figure}[!t]
\centering
\includegraphics[width=2.5in]{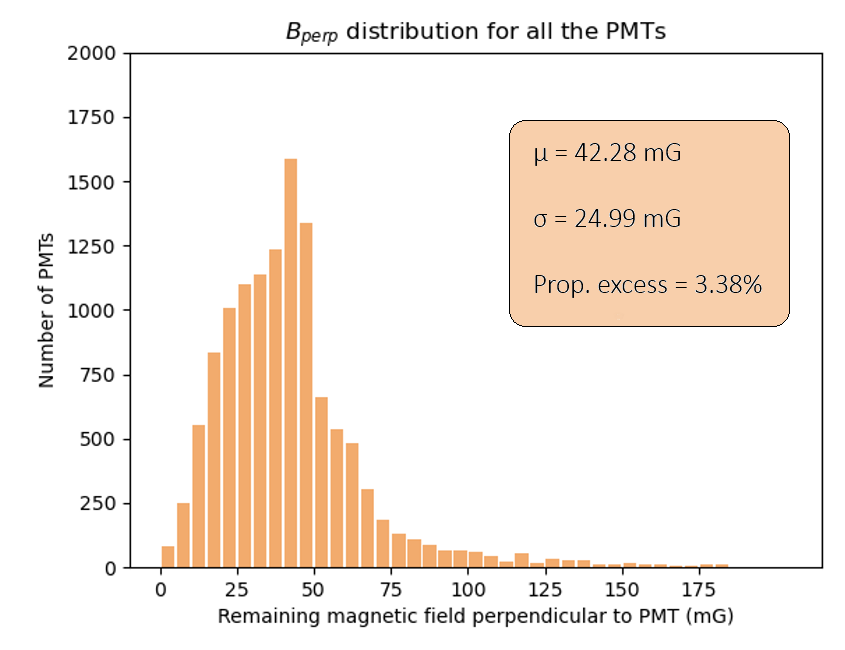}
\caption{Histogram of $<B_{perp}>$ after adding more turns to the top and bottom circular coils and elliptical coils to minimize $<B_{perp}>$.}
\label{Figure21}
\end{figure}

Elliptical coils are now simulated with current intensities optimized to minimize $<B_{perp}>$. The results are presented in Figure \ref{Figure21}. In this configuration, $P_{100}$ decreases to 3.38\%, although this reduction is less significant than when directly minimizing $P_{100}$. However, the average $<B_{perp}>$ decreases substantially to $42.28 \pm 24.99\: mG$, leading to an average efficiency loss of only $0.91 \pm 0.92\%$. This optimization also significantly reduces the proportion of PMTs experiencing greater than 1\% loss to 25.84\%, and the proportion experiencing greater than 5\% loss to 1.01\%.\\

\begin{figure}[!t]
\centering
\includegraphics[width=2.5in]{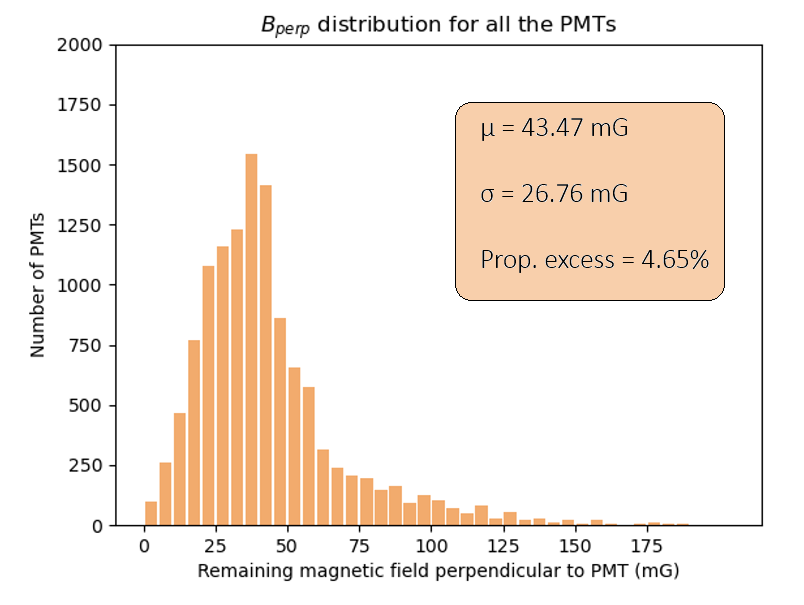}
\caption{Histogram of $<B_{perp}>$ after adding more turns to the top and bottom circular coils and smaller circular coils to minimize $<B_{perp}>$.}
\label{Figure22}
\end{figure}

If the proposed smaller-radius circular coils are used instead of elliptical coils, the results depicted in Figure \ref{Figure22} are obtained. In this case, optimization leads to radii of R = 15 m and R = 14 m for the upper and lower coils, respectively. The resulting $P_{100}$ value is 4.65\%, which is higher than that achieved with elliptical coils, confirming the superior performance of the latter for minimizing this parameter. The average $<B_{perp}>$, however, is $43.47 \pm 26.76\: mG$, exceeding that of the elliptical coil configuration by over one point. Consequently, the average efficiency loss is slightly higher at $0.95 \pm 0.98\%$. The proportions of PMTs experiencing losses greater than 1\% and 5\% are also slightly elevated, reaching 29.29\% and 0.99\%, respectively.\\

Therefore, elliptical coils prove more effective in compensating for both parameters due to their localized effect on areas where PMTs remain inadequately compensated. Nevertheless, the results obtained with smaller-radius circular coils, while slightly inferior, are sufficiently comparable, particularly when optimizing $<B_{perp}>$. This makes them a more practical alternative in terms of installation, cost (due to the reduced number of power supplies and comparable cable length), and overall compensation performance.\\

\section{Conclusions}\label{sec5}

Compensating for the Earth's magnetic field in a cylindrical Cherenkov detector using a coil-based system is complex due to the numerous factors involved, ranging from the detector's dimensions to the parameter selected for evaluating the compensation effectiveness. This paper examines the suitability of using the proportion of PMTs exposed to a perpendicular magnetic field exceeding 100 mG as a metric for assessing the geomagnetic field compensation. This approach is compared to minimizing the average total magnetic field ($<B_{total}>$) and the average perpendicular magnetic field ($<B_{perp}>$) by calculating the efficiency loss in each PMT caused by the residual magnetic field.\\

The results demonstrate that compensation effectiveness increases with the distance between the PMTs and the detector walls, irrespective of the evaluation parameter used. Furthermore, the optimal coil spacing also increases as the distance between the PMTs and the walls grows.\\

Concerning the choice of the parameter to minimize, $<B_{perp}>$ emerges as a suitable metric when the objective is to reduce the average detection efficiency loss and the number of PMTs affected. Conversely, minimizing the proportion of PMTs exceeding the 100 mG threshold proves most beneficial when specifically aiming to minimize the number of PMTs with significant efficiency losses. However, the difference between optimizing this parameter and the other two is less pronounced for this specific purpose.\\

It is also worth noting that, given the approximations used in this study, the parameters examined for coil optimization yield results that are generally quite comparable. The small differences observed make it challenging to define a clear preference for one evaluation metric over another. Nevertheless, these findings provide a robust framework for understanding the trade-offs associated with each optimization strategy.\\

Finally, the optimization of both parameters, $P_{100}$ and $<B_{perp}>$, in a concrete simulation example is compared, finding that optimizing $<B_{perp}>$ is more advantageous for reducing detection efficiency loss. We also evaluate the addition of elliptical coils at both ends against the installation of circular coils with equal radii, centered at both ends, to address areas where compensation is most challenging. While elliptical coils yield superior compensation results, the performance of smaller circular coils proves sufficiently similar and satisfactory. Consequently, these circular coils emerge as a more feasible and economical alternative for practical installation.\\

\bmhead{Acknowledgments}

This research paper was supported by Boosting Ingenium for Excellence (BI4E) project,funded by the European Union's European Union’s HORIZON-WIDERA-2021-ACCESS-05-01-European Excellence Initiative under the Grant Agreement No. 101071321. It has also been developed with financial support from the grant by the MRR-22-MCINN-HKK-CANFRANC project founded by MICINN and FICYT regional plans.  SLS, MLS and JDS acknowledge financial support from project PID2022-139198NB-I00. LB acknowledges financial support from the PID2021-125630NB-I00 project funded by MCIN/AEI/10.13039/501100011033 / FEDER, UE. FJDC acknowledges financial support from the PID2021-127331NB-I00 project funded by MCIN/AEI/10.13039/501100011033 / FEDER, UE.\\

\bmhead{Data Availability Statement}

The datasets generated during and/or analysed during the current study are available from the corresponding author on reasonable request.

\end{document}